\documentclass[journal]{IEEEtran}
\usepackage[latin1]{inputenc}
\usepackage{amsfonts}
\usepackage{amssymb}
\usepackage{graphicx}
\usepackage{verbatim}
\usepackage{array}
\usepackage{multirow}
\usepackage{dcolumn}
\usepackage{color}
\usepackage{flushend}
\usepackage{url}
\usepackage{stfloats}
\usepackage{siunitx}
\usepackage{rotating}

\usepackage[tbtags]{amsmath}

\usepackage{eurosym}

\usepackage{multirow}

\usepackage{color}

\def \R {{\rm I\kern -2.2pt R\hskip 1pt}}

\AtEndDocument{\par\leavevmode}

\DeclareGraphicsExtensions{.png} 

\begin{document}
\title{On the Solution of Large-Scale Robust Transmission Network Expansion Planning under Uncertain Demand and Generation Capacity}

\author{R. Mínguez,
        R. García-Bertrand,~\IEEEmembership{Senior Member,~IEEE,}
        José~M.~Arroyo,~\IEEEmembership{Senior Member,~IEEE,}
        and Natalia Alguacil,~\IEEEmembership{Senior Member,~IEEE}

\thanks{This work was supported in part by the Ministry of Science of Spain under CICYT Project ENE2015-63879-R (MINECO/FEDER, UE), the Junta de Comunidades de Castilla-La Mancha under Project POII-2014-012-P, and the Universidad de Castilla-La Mancha under Grant GI20163388.}

\thanks{R. Mínguez is with Hidralab Ingeniería y Desarrollo, S.L., Spin-Off UCLM, Hydraulics Laboratory, Universidad
 de Castilla-La Mancha, Ciudad Real E-13071, Spain (e-mail: \mbox{roberto.minguez@hidralab.com}).

R. García-Bertrand, J. M. Arroyo, and N. Alguacil are with the Departamento de Ingeniería Eléctrica,
Electrónica, Automática y Comunicaciones, E.T.S.I. Industriales, Universidad
 de Castilla-La Mancha, Ciudad Real E-13071, Spain (e-mail: \mbox{Raquel.Garcia@uclm.es}; \mbox{JoseManuel.Arroyo@uclm.es}; \mbox{Natalia.Alguacil@uclm.es}).

}}

\maketitle

\begin{abstract}
Two-stage robust optimization has emerged as a relevant approach to deal with uncertain demand and generation capacity in the transmission network expansion planning problem. Unfortunately, available solution methodologies for the resulting trilevel robust counterpart are unsuitable for large-scale problems. In order to overcome this shortcoming, this paper presents an alternative column-and-constraint generation algorithm wherein the max-min problem associated with the second stage is solved by a novel coordinate descent method. As a major salient feature, the proposed approach does not rely on the transformation of the second-stage problem to a single-level equivalent. As a consequence, bilinear terms involving dual variables or Lagrange multipliers do not arise, thereby precluding the use of computationally expensive big-M-based linearization schemes. Thus, not only is the computational effort reduced, but also the typically overlooked non-trivial tuning of bounding parameters for dual variables or Lagrange multipliers is avoided. The practical applicability of the proposed methodology is confirmed by numerical testing on several benchmarks including a case based on the Polish 2383-bus system, which is well beyond the capability of the robust methods available in the literature.
\end{abstract}


\section*{Nomenclature}

This section lists the main notation used throughout the paper.
Additional symbols with superscripts ``$(k)$'' and ``$(m)$'' are used
to indicate the value of a specific variable at iterations $k$ and $m$ of the column-and-constraint generation algorithm, respectively. Similarly, superscript ``$(\nu)$'' is used to denote results obtained at iteration $\nu$ of the coordinate descent method.


\subsection{Indices}
\begin{description}
[\setlabelwidth{$\mbox{\it{ELLL}}$}\usemathlabelsep]

\item[$i$] Generating unit index.

\item[$j$] Load index.

\item[$l$] Transmission line index.

\item[$n$] Bus index.




\end{description}

\subsection{Sets}
\begin{description}
[\setlabelwidth{$\mbox{\it{ELLL}}$}\usemathlabelsep]

\item[$\Omega^{\rm D}$] Set of indexes of loads.

\item[$\Omega^{\rm D}_n$] Set of indexes of loads connected to bus $n$.
\vspace{0.1cm}

\item[$\Omega^{\rm G}$] Set of indexes of generating units.

\item[$\Omega^{\rm G}_n$] Set of indexes of generating units connected to bus $n$.

\item[$\Omega^{\rm L}$] Set of indexes of existing transmission lines.

\item[$\Omega^{\rm L^{\rm +}}$] Set of indexes of candidate transmission lines.

\item[$\Omega^{\rm N}$] Set of indexes of buses.



\end{description}






\subsection{Constants}
\begin{description}
[\setlabelwidth{$\mbox{\it{ELLL}}$}\usemathlabelsep]

\item[$\gamma_{j}$] Fraction of the demand of load $j$ that can be curtailed.

\item[$\Gamma^{\rm D}$] Conservativeness parameter for uncertain demands.

\item[$\Gamma^{\rm G}$] Conservativeness parameter for uncertain generation capacities.



\item[$\Delta_{j}^{\rm D}$] Maximum deviation from the nominal or forecast demand of load $j$.

\item[$\Delta_{i}^{\rm G}$] Maximum deviation from the nominal or forecast generation capacity of unit $i$.

\item[$\epsilon^{\rm {IL}}$] Convergence tolerance for the inner loop.

\item[$\epsilon^{\rm {OL}}$] Convergence tolerance for the outer loop.

\item[$\Pi$] Investment budget.

\item[$\sigma$] Weighting factor.

\item[$C^{\rm L}_l$] Construction cost of candidate line $l$.
\vspace{0.1cm}

\item[$C^{\rm G}_{i}$] Production cost coefficient of unit $i$.
\vspace{0.1cm}

\item[$C^{\rm U}_{j}$] Load-shedding cost coefficient of load $j$.
\vspace{0.1cm}

\item[$\overline{F}_{l}$] Power flow capacity of line $l$.

\item[$fr(l)$] Sending or origin bus of line $l$.
\vspace{0.1cm}

\item[$p_{j}^{{\rm D},f}$] Nominal or forecast demand of load $j$.
\vspace{0.1cm}


\item[$\overline{p}_{i}^{{\rm G},f}$] Nominal or forecast generation capacity of unit $i$.
\vspace{0.1cm}


\item[$to(l)$] Receiving or destination bus of line $l$.

\item[$x_l$] Reactance of line $l$.










\end{description}


\subsection{Decision Variables}

\begin{description}
[\setlabelwidth{$\mbox{\it{ELLL}}$}\usemathlabelsep]

\item[$\alpha$] Approximation of the worst-case operating cost.

\item[$\theta_n$] Phase angle at bus $n$.

\item[$\theta_{nm}$] Phase angle at bus $n$ under the worst case identified at iteration $m$.

\item[$c^{\rm {IL}}$] Operating cost resulting from the inner loop.

\item[$c^{\rm O}$] Operating cost.

\item[$c^{{\rm O},cd}$] Operating cost computed along the iterations of the coordinate descent algorithm.

\item[$c^{{\rm O},wc}$] Worst-case operating cost.

\item[$c^{\rm {OL}}$] Total cost resulting from the outer loop.

\item[$p^{\rm D}_{j}$] Uncertain demand of load $j$.
\vspace{0.1cm}


\item[$p^{\rm G}_{i}$] Power output of unit $i$.
\vspace{0.1cm}

\item[$p^{\rm G}_{im}$] Power output of unit $i$ under the worst case identified at iteration $m$.
\vspace{0.1cm}

\item[$\overline p^{\rm G}_{i}$] Uncertain generation capacity of unit $i$.
\vspace{0.1cm}

\item[$p^{\rm L}_{l}$] Power flow through line $l$.
\vspace{0.1cm}

\item[$p^{\rm L}_{lm}$] Power flow through line $l$ under the worst case identified at iteration $m$.
\vspace{0.1cm}

\item[$p^{\rm U}_{j}$] Unserved demand of load $j$.
\vspace{0.1cm}

\item[$p^{\rm U}_{jm}$] Unserved demand of load $j$ under the worst case identified at iteration $m$.

\item[$v_l$] Binary variable that is equal to 1 if candidate line $l$ is built, being 0 otherwise.

\item[$z^{\rm D}_{j}$] Binary variable that is equal to 1 if the worst-case demand of load $j$  is equal to its
upper bound, being 0 otherwise.

\item[$z^{\rm G}_{i}$] Binary variable that is equal to 1 if the worst-case generation capacity of unit $i$  is equal to its
lower bound, being 0 otherwise.






\end{description}

\subsection{Dual Variables}
\begin{description}
[\setlabelwidth{$\mbox{\it{ELLL}}$}\usemathlabelsep]

\item[$\mu^{\rm D}_{j}$] Dual variable associated with the demand of load $j$.

\item[$\mu^{\rm G}_{i}$] Dual variable associated with the generation capacity of unit $i$.

\end{description}


\section{Introduction}


\IEEEPARstart{T}{ransmission} network expansion planning is a key decision-making problem under both non-competitive and market-based settings. The traditional transmission network expansion planning problem consists in determining the optimal investments in transmission facilities so that power is supplied to consumers in a reliable and economic fashion \cite{Wang:94}. The growing penetration levels of renewable-based generation have confronted network planners with major challenges \cite{HemmatiHK:13}, \cite{LumbrerasR:16}. First, the  uncertainty in the nodal net power injections, which was traditionally associated with demand growth, has drastically increased. In addition, the installation of renewable-based generation facilities becomes a relevant uncertain aspect itself. Moreover, as a major complicating factor, accurate probability distributions for such new sources of uncertainty are unavailable within a planning horizon.

Such challenges have triggered significant research effort to effectively address transmission network expansion planning under uncertain demand and generation capacity. Relevant approaches rely on the use of scenarios \cite{Reis:05}, \cite{Yu:11}, intervals \cite{Escobar:08}, and chance-constrained programming \cite{Yu:09}. In order to overcome the limitations of the methods described in \cite{Reis:05}--\cite{Yu:09}, recent contributions \cite{WuCX:08}--\cite{Minguez:16} suggest the use of two-stage adaptive or adjustable robust optimization (ARO) \cite{Ben-Tal:04}. Unlike scenario-based methods \cite{Reis:05}, \cite{Yu:11} and chance-constrained programming \cite{Yu:09}, ARO neither requires accurate probabilistic information nor relies on a discrete set of uncertainty realizations requiring a tradeoff between tractability and accuracy that may be hard to attain. Rather, uncertainty is modeled by decision variables within an uncertainty set, thereby comprising an infinite number of uncertainty realizations. Hence, the size of the robust counterpart does not depend on the dimension of the space of uncertainty realizations belonging to the uncertainty set, which is beneficial for implementation purposes. The uncertainty set can be built using intervals defined by the lower and upper bounds for the uncertain parameters. Such information, which is similar to that required by the interval-based method presented in \cite{Escobar:08}, may be easier to derive than probability distributions \cite{Ruiz:15}. Moreover, and in contrast to \cite{Escobar:08}, the robust solution guards against all realizations of uncertainty within the uncertainty set. Such a worst-case setting is a particularly desirable feature in planning problems \cite{ChenWWHW:14}, \cite{Ruiz:15}.

Within an ARO-based setting \cite{WuCX:08}--\cite{Minguez:16}, the robust counterpart is formulated as an instance of trilevel programming wherein the first stage is associated with the upper level and the second stage corresponds to the max-min problem characterizing the two lowermost optimization levels. The upper level determines the least-cost first-stage decisions, namely the optimal investment plan. For a given upper-level decision vector, the middle level identifies the worst-case values of uncertain demand and generation capacity leading to maximum operating cost. Finally, the lower level models the operator's best reaction, by means of adjustable variables, that minimizes the operating cost for given upper- and middle-level decisions.

In \cite{WuCX:08}, the resulting trilevel program was addressed by a greedy randomized adaptive search
procedure combined with a modified branch-and-bound algorithm for the second-stage problem. Alternative approaches applied Benders decomposition \cite{Jabr:13} and the column-and-constraint generation algorithm \cite{ChenWWHW:14}--\cite{Minguez:16}, both involving the iterative solution of a master problem and a max-min subproblem associated with the second stage. However, available methods feature limitations.

Similar to the heuristic applied in \cite{WuCX:08}, the approaches presented in \cite{Jabr:13}--\cite{Minguez:16} are unable to acknowledge global optimality. This shortcoming results from the transformation of the subproblem into a mixed-integer linear equivalent relying on setting bounds for dual variables or Lagrange multipliers, which may be in general unbounded \cite{Luenberger:84}. Such unboundedness precludes the development of effective selection procedures for those bounds, being heuristic trial-and-error algorithms \cite{Ruiz:15} the current strategy.  Thus, there is no guarantee that the subproblem is solved to optimality, which is the requirement for the exactness of the above-mentioned decomposition-based methods.

Moreover, most of the literature in this area has focused on showing the benefits of using robust optimization to deal with uncertainty, while acknowledging that there are still computational hurdles to be overcome when solving large real-life instances. In theory, existing robust approaches \cite{WuCX:08}--\cite{Minguez:16} can be successfully implemented through the use of off-the-shelf software. As reported in \cite{Jabr:13}, \cite{Minguez:16}, however, only relatively small instances appear to be tractable with such approaches, being the solution described in \cite{Minguez:16} the most efficient. Such intractability results from the transformation of the max-min second stage to an equivalent single-level albeit bilinear problem comprising highly nonconvex products of binary variables and dual variables or Lagrange multipliers, for which the specific tools devised in \cite{WuCX:08}--\cite{Minguez:16} are impractical. Thus, the intractability issue remains a challenging obstacle to the practical implementation of the methods presented in \cite{WuCX:08}--\cite{Minguez:16}.

Motivated by both facts, the thrust of this paper is the proposal of a novel and computationally efficient methodology suitable for large-scale instances of the robust transmission network expansion planning problem under uncertain demand and generation capacity addressed in \cite{Jabr:13}--\cite{Minguez:16}. The proposed approach is a modified column-and-constraint generation algorithm solely involving primal decision variables wherein the max-min subproblem is solved by a novel coordinate descent algorithm particularly tailored to its bilevel structure. Thus, as a distinctive feature over previous duality-based methods \cite{WuCX:08}--\cite{Minguez:16}, which require the iterative and time consuming solution of a large-scale mixed-integer bilinear subproblem or its linearized equivalent, two simpler primal problems respectively associated with the middle- and lower-level problems are iteratively solved. As a consequence, the computational burden is substantially decreased since neither duality-based cuts, nor a single-level transformation based on dual variables or Lagrange multipliers, nor a linearization scheme, nor a case-dependent, non-trivial, and computationally expensive bounding parameter selection procedure are required. Although, similar to previous works \cite{WuCX:08}--\cite{Minguez:16}, global optimality is not guaranteed,  the proposed approach is capable of attaining high-quality near-optimal investment decisions for large-scale instances that are unsolvable by available methods \cite{WuCX:08}--\cite{Minguez:16}.





The main contributions of this paper are twofold:

\begin{enumerate}
    \item From a methodological perspective, this paper presents a new primal column-and-constraint generation algorithm for robust transmission network expansion planning under uncertainty that allows effectively dealing with real-life instances of this problem.
    \item For the first time in the literature on robust transmission network expansion planning under uncertainty, this paper reports successful numerical experience with a large-scale test system comprising thousands of buses and lines, which is well beyond the capability of existing approaches.
\end{enumerate}

The remainder of the paper is structured as follows. The ARO-based framework for robust transmission network expansion planning under uncertainty is described in Section \ref{rtnep}, wherein both the uncertainty characterization and the formulation of the robust counterpart are provided. Section \ref{ccg} is devoted to the proposed solution approach. In Section \ref{CaseStudy}, numerical results are presented and analyzed. Finally, relevant conclusions are drawn in Section \ref{Conclusions}.

\section{Robust Transmission Network Expansion Planning}\label{rtnep}

The proposed application of adjustable or two-stage robust optimization \cite{Ben-Tal:04} relies on the uncertainty set and the robust counterpart described next.

\subsection{Uncertainty Characterization}\label{unc_char}

Under an ARO-based framework, uncertainty sources are characterized by the extremes of the respective fluctuation range. Such uncertainty characterization can be implemented by modeling uncertain demands and uncertain generation capacities as variables that can vary around their nominal values. The information on the uncertainty sources thus reduces to nominal values and fluctuation bounds. In addition, the conservativeness of this uncertainty modeling can be controlled by two user-defined parameters, denoted by uncertainty budgets or conservativeness parameters. Here, based on \cite{Jabr:13} and \cite{Minguez:16}, the uncertainty budgets represent the maximum numbers of demands and generating units that simultaneously experience fluctuations, respectively. The resulting cardinality uncertainty set can be mathematically cast as:
\begin{align}
&p_{j}^{{\rm D},f}-\Delta_{j}^{\rm D} \le p^{\rm D}_{j} \le p_{j}^{{\rm D},f}+\Delta_{j}^{\rm D}; \forall j \in {\Omega^{\rm D}} \label{intD}\\
&\overline{p}_{i}^{{\rm G},f}-\Delta_{i}^{\rm G} \le \overline{p}^{\rm G}_{i} \le \overline{p}_{i}^{{\rm G},f}+\Delta_{i}^{\rm G}; \forall i \in {\Omega^{\rm G}} \label{intG}\\
&\sum_{j \in {\Omega^{\rm D}} }\Biggr \lceil{\frac{\lvert p^{ \rm D}_{j} - p_{j}^{{\rm D},f}\rvert\  }{\Delta_{j}^{\rm D}}\Biggr\rceil }     \leq \Gamma^{\rm D}\label{intbudgetD}\\
&\sum_{i \in {\Omega^{\rm G}} }\Biggr \lceil{\frac{\lvert \overline{p}^{\rm G}_{i} - \overline{p}_{i}^{{\rm G},f}\rvert\  }{\Delta_{i}^{\rm G}}\Biggr\rceil }     \leq \Gamma^{\rm G}\label{intbudgetG}
\end{align}

\noindent where (\ref{intD}) and (\ref{intG}) model the demand- and generation-related intervals, respectively, whereas (\ref{intbudgetD}) and (\ref{intbudgetG}) respectively impose the uncertainty budgets for demand and generation capacity fluctuations.

In adjustable robust optimization with an uncertainty set defined by the fluctuation range of the parameters allowed to vary, the optimal values of the decision variables characterizing the uncertainty set, i.e., the worst-case values, are one of the extremes of the corresponding range. A proof can be found in \cite{WuCX:08}, where adjustable robust optimization was applied to the transmission network expansion planning problem under uncertain demand. Moreover, according to \cite{Ruiz:15}, the worst case corresponds to generation capacities being as low as possible and demands being as high as possible. Thus, under the above-defined demand uncertainty budget, the worst-case demands, which represent all possible realizations of uncertain demands, can take two values, namely the upper bound $p_{j}^{{\rm D},f}+\Delta_{j}^{\rm D}$, or the forecast value $p_{j}^{{\rm D},f}$. As a consequence, for each uncertain demand, all possible uncertainty realizations can be modeled by a binary variable. Analogously, under the above-defined generation uncertainty budget, the worst-case generation capacities, which represent all possible realizations of uncertain generation capacities, can take two values, namely the lower bound $\overline{p}_{i}^{{\rm G},f}-\Delta_{i}^{\rm G}$, or the forecast value $\overline{p}_{i}^{{\rm G},f}$. Hence, for each uncertain generation capacity, all possible uncertainty realizations can be modeled by an additional binary variable.

Thus, as done in \cite{Minguez:16}, the uncertainty set adopted here is formulated as follows:
\begin{align}
&p^{\rm D}_{j}=p_{j}^{{\rm D},f}+\Delta_{j}^{\rm D}z_{j}^{\rm D}; \forall j \in {\Omega^{\rm D}} \label{worstD}\\
&\overline{p}^{\rm G}_{i}=\overline{p}_{i}^{{\rm G},f}-\Delta_{i}^{\rm G}z_{i}^{\rm G}; \forall i \in {\Omega^{\rm G}} \label{worstG}\\
&z_{j}^{\rm D} \in \{0,1\}; \forall j \in {\Omega^{\rm D}} \label{binary_D}\\
&z_{i}^{\rm G} \in \{0,1\}; \forall i \in {\Omega^{\rm G}} \label{binary_G}\\
&\sum_{j \in {\Omega^{\rm D}} }z_{j}^{\rm D} \leq \Gamma^{\rm D}\label{budgetD}\\
&\sum_{i \in {\Omega^{\rm G}} }z_{i}^{\rm G} \leq \Gamma^{\rm G}.\label{budgetG}
\end{align}

Constraints (\ref{worstD}) and (\ref{worstG}) respectively express the worst-case levels of demand and generation capacity in terms of the corresponding nominal levels and fluctuation levels. To that end, binary variables $z_{j}^{\rm D}$ and $z_{i}^{\rm G}$ are used, the integrality of which is modeled in (\ref{binary_D}) and (\ref{binary_G}), respectively. The conservativeness of the uncertainty characterization is modeled in (\ref{budgetD}) and (\ref{budgetG}), where demand- and generation-related uncertainty budgets are respectively imposed.

\subsection{Problem Formulation}


Based on the static models presented in \cite{ChenWWHW:14} and \cite{Ruiz:15}, the two-stage adaptive robust optimization model for the transmission network expansion planning problem under uncertainty can be formulated as the following mixed-integer trilevel program:
\begin{align}
&\underset{{c^{{\rm O},wc}},v_{l}}{\text{Minimize}} \sum_{l\in \Omega^{\rm L^{\rm +}}}C^{\rm L}_{l}v_{l} +\sigma{c^{{\rm O},wc}}\label{of1}\\
&\text{subject to:}\notag\\
& \sum_{l\in \Omega^{\rm L^{\rm +}}    }C^{\rm L}_{l}v_{l} \leq \Pi \label{invest max}\\
& v_{l} \in \{0,1\}; \forall l \in \Omega^{\rm L^{\rm +}}\label{decision var}\\
&c^{{\rm O},wc}=\underset{{c^{{\rm O}}},p^{\rm D}_{j},\overline{p}^{\rm G}_{i},z^{\rm D}_{j},z^{\rm G}_{i}}{\text{Maximize}} \quad \Bigg \{ {c^{{\rm O}}}\label{of2}\\
&\quad \text{subject to:}\notag\\
&\quad\text{Constraints (\ref{worstD})--(\ref{budgetG})}\label{compact}\\
&\quad{c^{{\rm O}}}=\underset{\theta_{n},p^{\rm G}_{i},p^{\rm L}_{l},p^{\rm U}_{j}}{\text{Minimize}}  \quad  \sum_{i\in \Omega^{\rm G}}C^{\rm G}_{i}p^{\rm G}_{i}+\sum_{j\in \Omega^{\rm D}}C^{\rm U}_{j}p^{\rm U}_{j}\label{of3}\\
&\quad\hspace{0.3cm} \text{subject to:}\notag\\
&\quad\hspace{0.3cm} \sum_{i \in \Omega_n^{\rm G}}{p^{\rm G}_{i}}+\sum_{j \in {\Omega_n^{\rm D}}}{p^{\rm U}_{j}}+\sum_{l \in (\Omega^{\rm L}    \cup  \Omega^{\rm L^{\rm +}} )| {to}(l)=n}{p^{\rm L}_{l}} \nonumber\\
&\quad\hspace{0.3cm} - \sum_{l \in (\Omega^{\rm L}    \cup  \Omega^{\rm L^{\rm +}} )| {fr}(l)=n}{p^{\rm L}_{l}}= \sum_{j \in {\Omega_n^{\rm D}}}{p_{j}^{\rm D}};\forall n \in \Omega^{\rm N}\label{balance}\\
&\quad\hspace{0.3cm} p^{\rm L}_{l}=\frac{1}{x_l}(\theta_{fr(l)}-\theta_{to(l)}); \forall l \in \Omega^{\rm L}\label{flow_existing}\\
&\quad\hspace{0.3cm} p^{\rm L}_{l}=\frac{v_{l}}{x_l}(\theta_{fr(l)}-\theta_{to(l)}); \forall l \in \Omega^{\rm L^{\rm +}}\label{flow}\\
&\quad\hspace{0.3cm}-\overline{F}_{l}\leq p^{\rm L}_{l}\leq \overline{F}_{l}; \forall l \in (\Omega^{\rm L}    \cup  \Omega^{\rm L^{\rm +}} )\label{flow_lim}\\
&\quad\hspace{0.3cm} 0 \leq p^{\rm G}_{i} \leq \overline{p}^{\rm G}_{i};\forall i \in {\Omega^{\rm G}}\label{gen_lim}
\end{align}
\begin{align}
&\quad\hspace{0.3cm} 0 \leq p^{\rm U}_{j} \leq \gamma_jp^{\rm D}_{j};\forall j \in {\Omega^{\rm D}}.\label{shed_lim}
\end{align}

Problem (\ref{of1})--(\ref{shed_lim}) is a trilevel program comprising three optimization
levels: 1) the upper level (\ref{of1})--(\ref{decision var}), which is associated with the identification of the least-cost expansion plan; 2) the middle level (\ref{of2})--(\ref{compact}), characterizing the worst-case realization of uncertainty sources for a given upper-level investment plan; and 3) the lower level (\ref{of3})--(\ref{shed_lim}), which is related to the optimal system operation for given upper-level investment decisions and middle-level uncertainty realizations.

The objective of the upper-level problem is the minimization of the total cost (\ref{of1}), which comprises two terms, namely the annualized investment cost and the worst-case operating cost. The weighting factor $\sigma$ is used to make investment and worst-case operating costs comparable quantities. The upper-level minimization is subject to an upper bound on the investment cost (\ref{invest max}). In addition, the binary nature of investment variables is modeled in (\ref{decision var}).

The middle-level problem (\ref{of2})--(\ref{compact}) identifies the worst-case uncertainty realizations yielding the largest operating cost (\ref{of2}) for the solution identified by the upper level. The uncertainty characterization described in Section \ref{unc_char} is modeled by (\ref{compact}).

In the lower-level problem (\ref{of3})--(\ref{shed_lim}), the operating cost associated with upper- and middle-level
variables is minimized in (\ref{of3}). Expressions (\ref{balance})--(\ref{flow_lim}) model the effect of the network including nodal power balances  (\ref{balance}), line flows through existing lines (\ref{flow_existing}), line flows through candidate lines (\ref{flow}), and line flow limits (\ref{flow_lim}). Constraints (\ref{gen_lim}) set the generation limits. Finally, constraints (\ref{shed_lim}) impose bounds on load shedding.

\section{Solution Approach}\label{ccg}

The proposed solution approach is a primal column-and-constraint generation algorithm, hereinafter referred to as P-CC. Unlike previous applications of the column-and-constraint generation algorithm \cite{ChenWWHW:14}--\cite{Minguez:16}, which require dealing with highly nonconvex bilinear terms including dual lower-level variables, P-CC solely relies on linear expressions involving primal decision variables, which is beneficial for computational purposes. The master problem and the subproblem that are solved along the iterations are described next followed by an outline of the proposed iterative process.

\subsection{Master Problem}

The master problem constitutes a relaxation for problem (\ref{of1})--(\ref{shed_lim}) where a set of valid operating constraints are iteratively added. The addition of such constraints, which are set up with information from the subproblem, allows obtaining a more robust expansion plan at each iteration. At iteration $k$, the master problem is formulated as the following mixed-integer linear program:
\begin{align}
&\hspace{-0.2cm}\underset{\alpha,\theta_{nm},p^{\rm G}_{im},p_{lm}^{\rm L},p^{\rm U}_{jm},v_{l}}{\text{Minimize}}\hspace{0.2cm} \sum_{l\in \Omega^{\rm L^{\rm +}}}C^{\rm L}_{l}v_{l} +\sigma\alpha\label{mof1}\\
&\hspace{-0.2cm}\text{subject to:}\notag\\
&\hspace{-0.2cm}\text{Constraints (\ref{invest max}) and (\ref{decision var})}\label{mcompact}\\
&\hspace{-0.2cm}\alpha \geq \sum_{i\in \Omega^{\rm G}}C^{\rm G}_{i}p^{\rm G}_{im}+\sum_{j\in \Omega^{\rm D}}C^{\rm U}_{j}p^{\rm U}_{jm}; m=1,\ldots, k-1  \label{mof3}
\end{align}
\begin{align}
&\hspace{-0.2cm} \sum_{i \in \Omega_n^{\rm G}}{p^{\rm G}_{im}}+\sum_{j \in {\Omega_n^{\rm D}}}{p^{\rm U}_{jm}}+\sum_{l \in (\Omega^{\rm L}    \cup  \Omega^{\rm L^{\rm +}} )| {to}(l)=n}{p^{\rm L}_{lm}} \nonumber\\
&\hspace{-0.2cm} - \sum_{l \in (\Omega^{\rm L}    \cup  \Omega^{\rm L^{\rm +}} )| {fr}(l)=n}{p^{\rm L}_{lm}}= \sum_{j \in {\Omega_n^{\rm D}}}{p_{j}^{{\rm D}(m)}};\forall n \in \Omega^{\rm N}, \nonumber\\
&\hspace{-0.2cm} m=1,\ldots, k-1\label{mbalance}\\
&\hspace{-0.2cm} p_{lm}^{\rm  L}=\frac{1}{x_l}(\theta_{fr(l)m}-\theta_{to(l)m}); \forall l \in \Omega^{\rm L}, \nonumber\\
&\hspace{-0.2cm} m=1,\ldots, k-1\label{mflow_existing}\\
&\hspace{-0.2cm} p_{lm}^{\rm L}=\frac{v_{l}}{x_l}(\theta_{fr(l)m}-\theta_{to(l)m}); \forall l \in \Omega^{\rm L^{\rm +}}, \nonumber\\
&\hspace{-0.2cm} m=1,\ldots, k-1 \label{mflow}\\
&\hspace{-0.2cm} -\overline{F}_{l}\leq p_{lm}^{\rm L}\leq \overline{F}_{l}; \forall l \in (\Omega^{\rm L}    \cup  \Omega^{\rm L^{\rm +}} ), m=1,\ldots, k-1\label{mflow_lim}\\
&\hspace{-0.2cm} 0 \leq p^{\rm G}_{im} \leq \overline{p}^{{\rm G}(m)}_{i};\forall i \in {\Omega^{\rm G}}, m=1,\ldots, k-1\label{mgen_lim}\\
&\hspace{-0.2cm} 0 \leq p^{\rm U}_{jm} \leq \gamma_jp^{{\rm D}(m)}_{j};\forall j \in {\Omega^{\rm D}}, m=1,\ldots, k-1\label{mshed_lim}\\
&\hspace{-0.2cm} \alpha  \geq 0 \label{neg}
\end{align}
\noindent where the additional decision variables, $\theta_{nm}$, $p^{\rm G}_{im}$, $p_{lm}^{\rm L}$, and $p^{\rm U}_{jm}$, corresponding to $\theta_{n}$, $p^{\rm G}_{i}$, $p^{\rm L}_{l}$, and $p^{\rm U}_{j}$, respectively, are associated with the demands and generation capacities identified by the subproblem at iteration $m$ through $p^{{\rm D}(m)}_{j}$ and $\overline{p}^{{\rm G}(m)}_{i}$.

The objective function (\ref{mof1}) is identical to (\ref{of1}) except for the last term, where ${c^{{\rm O},wc}}$ is replaced with $\alpha$, which represents the pointwise maximum within all linear approximations of ${c^{{\rm O},wc}}$. Expression (\ref{mcompact}) includes the upper-level constraints. As modeled in (\ref{mof3}), the operating cost corresponding to the uncertainty realizations identified at iteration $m$ represents a lower bound for $\alpha$. Constraints (\ref{mbalance})--(\ref{mshed_lim}) respectively correspond to lower-level constraints (\ref{balance})--(\ref{shed_lim}). Finally, the nonnegativity of $\alpha$ is imposed in (\ref{neg}).

\subsection{Subproblem}

At each iteration $k$, the subproblem determines the worst-case uncertainty realizations yielding the maximum operating cost for a given upper-level decision provided by the previous master problem. Mathematically, the subproblem is a mixed-integer linear max-min problem comprising the two lowermost optimization levels (\ref{of2})--(\ref{shed_lim}) parameterized in terms of the given upper-level decision variables $v_l^{(k)}$.

Here we propose solving such bilevel problem through a block coordinate descent method \cite{Conejo:06} involving the iterative solution of two simple optimization problems. At iteration $\nu$, both problems are formulated as follows:
\begin{align}
&{c^{{\rm O},cd}}=\underset{\theta_{n},p^{\rm D}_{j},p^{\rm G}_{i},\overline{p}^{\rm G}_{i},p^{\rm L}_{l},p^{\rm U}_{j}}{\text{Minimize}}  \quad  \sum_{i\in \Omega^{\rm G}}C^{\rm G}_{i}p^{\rm G}_{i}+\sum_{j\in \Omega^{\rm D}}C^{\rm U}_{j}p^{\rm U}_{j}\label{of31}\\
&\hspace{0.3cm} \text{subject to:}\notag\\
&\hspace{0.3cm} \sum_{i \in \Omega_n^{\rm G}}{p^{\rm G}_{i}}+\sum_{j \in {\Omega_n^{\rm D}}}{p^{\rm U}_{j}}+\sum_{l \in (\Omega^{\rm L}    \cup  \Omega^{\rm L^{\rm +}} )| {to}(l)=n}{p^{\rm L}_{l}} \nonumber\\
&\hspace{0.3cm} - \sum_{l \in (\Omega^{\rm L}    \cup  \Omega^{\rm L^{\rm +}} )| {fr}(l)=n}{p^{\rm L}_{l}}= \sum_{j \in {\Omega_n^{\rm D}}}{p_{j}^{\rm D}};\forall n \in \Omega^{\rm N}\label{balance1}\\
&\hspace{0.3cm} p_{l}^{\rm L}=\frac{1}{x_l}(\theta_{fr(l)}-\theta_{to(l)}); \forall l \in \Omega^{\rm L}\label{flow_existing1}
\end{align}
\begin{align}
&\hspace{0.3cm} p_{l}^{\rm L}=\frac{v_{l}^{(k)}}{x_l}(\theta_{fr(l)}-\theta_{to(l)}); \forall l \in \Omega^{\rm L^{\rm +}}\label{flow1}\\
&\hspace{0.3cm}-\overline{F}_{l}\leq p_{l}^{\rm L}\leq \overline{F}_{l}; \forall l \in (\Omega^{\rm L}    \cup  \Omega^{\rm L^{\rm +}} )\label{flow_lim1}\\
&\hspace{0.3cm} 0 \leq p^{\rm G}_{i} \leq \overline{p}^{\rm G}_{i};\forall i \in {\Omega^{\rm G}}\label{gen_lim1}\\
&\hspace{0.3cm} 0 \leq p^{\rm U}_{j} \leq \gamma_jp^{\rm D}_{j};\forall j \in {\Omega^{\rm D}}\label{shed_lim1}\\
&\hspace{0.3cm} p^{\rm D}_{j} = p^{\rm D(\nu)}_{j}:\mu^{\rm D}_{j};\forall j \in {\Omega^{\rm D}}\label{sen_D}\\
&\hspace{0.3cm} \overline{p}^{\rm G}_{i}=\overline{p}^{\rm G(\nu)}_{i}:\mu^{\rm G}_{i};\forall i \in {\Omega^{\rm G}}\label{sen_G}
\end{align}
and
\begin{align}
&\underset{p^{\rm D}_{j},\overline{p}^{\rm G}_{i},z^{\rm D}_{j},z^{\rm G}_{i}}{\text{Maximize}} \quad  {c^{{\rm O},cd(\nu-1)}} + \sum_{i\in \Omega^{\rm G}}\mu^{\rm G(\nu-1)}_{i}(\overline{p}^{\rm G}_{i}-\overline{p}^{\rm G(\nu-1)}_{i})\nonumber\\
&\hspace{3cm}+\sum_{j\in \Omega^{\rm D}}\mu^{\rm D(\nu-1)}_{j}(p^{\rm D}_{j}- p^{\rm D(\nu-1)}_{j})   \label{of21}\\
&\text{subject to:}\notag\\
&\text{Constraints (\ref{worstD})--(\ref{budgetG})}\label{compact1}
\end{align}
\noindent where dual variables $\mu^{\rm D}_{j}$ and $\mu^{\rm G}_{i}$ represent the sensitivities of the operating cost with respect to fixed values of demands and generation capacities, respectively.

Problem (\ref{of31})--(\ref{sen_G}) corresponds to the lower-level problem (\ref{of3})--(\ref{shed_lim}) for fixed values of middle-level variables $p^{\rm D}_{j}$ and $\overline{p}^{\rm G}_{i}$. Such values of middle-level variables result from the optimal solution to problem (\ref{of21})--(\ref{compact1}), which corresponds to the middle-level problem (\ref{of2})--(\ref{compact}). Note that the operating cost in (\ref{of2}) is replaced in (\ref{of21}) with its first-order Taylor series approximation around the uncertainty realizations identified at the previous iteration of the coordinate descent algorithm. Those terms are based on the sensitivities $\mu^{\rm D}_{j}$ and $\mu^{\rm G}_{i}$ and the optimal operating cost ${c^{{\rm O},cd}}$ previously obtained from (\ref{of31})--(\ref{sen_G}).

Once initial values of middle-level variables $p^{\rm D}_{j}$ and $\overline{p}^{\rm G}_{i}$ are selected, the coordinate descent method iteratively solves problems (\ref{of31})--(\ref{sen_G}) and (\ref{of21})--(\ref{compact1}). The iterative process terminates when the operating cost remains unchanged within a user-defined tolerance. Admittedly, under the nonconvexity of the max-min subproblem, the algorithm may converge to a local optimum and hence the proposed column-and-constraint generation algorithm does not guarantee global convergence to optimality. Nonetheless, our expectation was that the proposed approach would still be a useful heuristic for finding good solutions. This was borne out by our computational experience, as described in Section~\ref{CaseStudy}.

\subsection{Algorithm}\label{Algorithm}

The proposed P-CC comprises two nested loops, namely 1) an outer loop associated with the master-subproblem iterations of the modified column-and-constraint generation algorithm, and 2) an inner loop related to the iterative process of the coordinate descent algorithm. The proposed methodology works as follows:

\begin{enumerate}
    \item \textit{Initialization of the outer loop.}
            \begin{itemize}
                \item Set the iteration counter of the outer loop $k$ to 1.
                \item Set the initial expansion plan $v_l^{(k)}=0$, $\forall l \in \Omega^{\rm L^{\rm +}}$ and the total cost associated with the outer loop $c^{\rm {OL}}$ to 0.
            \end{itemize}



    \item \label{initial_descent} \textit{Initialization of the inner loop.} Set the iteration counter of the inner loop $\nu$ to 1, select initial values for $p^{\rm D(\nu)}_{j}$ and $\overline{p}^{\rm G(\nu)}_{i}$, and set the operating cost associated with the inner loop $c^{\rm {IL}}$ to $+\infty$.

    \item \label{sub_step1} \textit{Solution of problem (\ref{of31})--(\ref{sen_G}).} Solve problem (\ref{of31})--(\ref{sen_G}) for the given expansion plan $v_l^{(k)}$ and given $p^{\rm D(\nu)}_{j}$ and $\overline{p}^{\rm G(\nu)}_{i}$. This step provides $\mu^{\rm D(\nu)}_{j}$, $\mu^{\rm G(\nu)}_{i}$, and ${c^{{\rm O},cd(\nu)}}$.

    \item \label{solution_check} \textit{Inner loop stopping criterion.} If a solution with a level of accuracy $\epsilon^{\rm {IL}}$ has been found, i.e., $|c^{\rm {IL}}-{c^{{\rm O},cd(\nu)}}|/|c^{\rm {IL}}| \leq \epsilon^{\rm {IL}}$, then go to step \ref{update_counters}; otherwise, set $c^{\rm {IL}} = c^{{\rm O},cd(\nu)}$ and go to step \ref{update_counters_inner}.

    \item \label{update_counters_inner} \textit{Update the iteration counter of the inner loop.} Increase the iteration counter $\nu \leftarrow \nu+1$.




    \item \label{sub_step2} \textit{Solution of problem (\ref{of21})--(\ref{compact1}).} Solve problem (\ref{of21})--(\ref{compact1}) for given $p^{\rm D(\nu-1)}_{j}$, $\overline{p}^{\rm G(\nu-1)}_{i}$, $\mu^{\rm D(\nu-1)}_{j}$, and $\mu^{\rm G(\nu-1)}_{i}$. This step provides $p^{\rm D(\nu)}_{j}$ and $\overline{p}^{\rm G(\nu)}_{i}$. Go to step \ref{sub_step1}.

    \item \label{update_counters} \textit{Update the iteration counter of the outer loop.} Increase the iteration counter $k \leftarrow k+1$.

    \item \label{step_master} \textit{Master problem solution.} Solve the master problem (\ref{mof1})--(\ref{neg}). This step provides $v_l^{(k)}$, $\alpha$, and the total cost of the outer loop, $c^{\rm {OL}} = \sum_{l\in \Omega^{\rm L^{\rm +}}}C^{\rm L}_{l}v_{l}^{(k)} +\sigma\alpha$.




    \item \label{stop_step} \textit{Outer loop stopping criterion}. If a solution with a level of accuracy $\epsilon^{\rm {OL}}$ has been found, i.e., $|\sum_{l\in \Omega^{\rm L^{\rm +}}}C^{\rm L}_{l}v_{l}^{(k)}+\sigma{c^{{\rm O},cd(\nu)}} -c^{\rm {OL}}|/|c^{\rm {OL}}| \leq \epsilon^{\rm {OL}}$, the algorithm stops; otherwise, go to step \ref{initial_descent}.

\end{enumerate}


\section{Numerical Results}\label{CaseStudy}

The performance of the proposed approach is illustrated with three cases respectively based on the IEEE 24-bus Reliability Test System (RTS)
\cite{RTS:99}, the IEEE 118-bus system \cite{118}, and the Polish 2383-bus system \cite{Polish}. For the sake of reproducibility, input data for all case studies can be downloaded from \cite{data}.

Numerical testing has been conducted for different values of the uncertainty budgets $\Gamma^{\rm D}$ and $\Gamma^{\rm G}$ including the case of maximum uncertainty wherein all uncertain demands and generation capacities are allowed to simultaneously fluctuate. Results from the proposed approach P-CC have been compared with those provided by the duality-based column-and-constraint generation algorithm presented in \cite{Minguez:16}, which, for quick reference, is hereinafter denoted by D-CC. Note that, to the best of the authors' knowledge, D-CC is the most computationally efficient method available in the literature on robust transmission network expansion planning under uncertainty. Simulations have been implemented on a Dell PowerEdge R920X64 with four Intel Xeon E7-4820 processors at 2.00 GHz and 768 GB of RAM using CPLEX 12.6 under GAMS 24.2 \cite{gams}. For all simulations, the optimality tolerance for the branch-and-cut algorithm of CPLEX was set at 10$^{-8}$. In addition, $\epsilon^{\rm {IL}}$ was set at 10$^{-12}$ whereas $\epsilon^{\rm {OL}}$, which was also used as the convergence tolerance for D-CC, was set at 10$^{-6}$.

\subsection{RTS-Based Case}
The first case study is based on the modified version of the IEEE RTS analyzed in \cite{Minguez:16}. This test system comprises 24 buses, 10 generating units, 17 loads, 34 existing lines, and 85 candidate lines \cite{data}. For illustration purposes, all demands and generation capacities are considered sources of uncertainty, which are respectively characterized by a $\pm$20$\%$ and a $\pm$50$\%$ fluctuation with respect to the corresponding nominal level.

Table \ref{single_24RTS} summarizes the best results provided by the proposed P-CC and D-CC for a \$20-million investment budget, for which the deterministic solution costs \$219161.7 million. For all instances, both approaches converged after 3  iterations of the outer loop in less than 2.2 s, thereby behaving similarly from a computational perspective. As for solution quality, the proposed P-CC attained the best total cost identified by D-CC for all instances but one corresponding to $\Gamma^{\rm D}= 5$ and $\Gamma^{\rm G}=3$. For such an instance, P-CC provided a lower bound for the optimal total cost that slightly differed by 0.0035\% from that achieved by D-CC. It is worth noting, however, that the best expansion decisions identified by both methods were identical for all combinations of uncertainty budgets. These results corroborate the effectiveness of the proposed approach to attain high-quality near-optimal solutions.

\begin{table}[t]
    \centering
    \caption{RTS-Based Case -- Results}
\begin{tabular}{ @{\extracolsep{3pt}}c@{\extracolsep{4pt}}c@{\extracolsep{5pt}}c@{\extracolsep{3pt}}c@{\extracolsep{5pt}}c@{\extracolsep{3pt}}c }
\cline{3-6}\noalign{\smallskip}&&\multicolumn{2}{c}{P-CC}&\multicolumn{2}{c}{D-CC}\\
\hline\noalign{\smallskip}
%
&&Total&Computing&Total&Computing\\
$\Gamma^{\rm D}$&$\Gamma^{\rm G}$&cost&time&cost&time\\
&&(10$^6$ \$)&(s)&(10$^6$ \$)&(s)\\\hline\noalign{\smallskip}
%
%
\textcolor[rgb]{1.00,1.00,1.00}{0}5&\textcolor[rgb]{1.00,1.00,1.00}{0}3&348929.5&0.9&348941.6&1.2\\
\textcolor[rgb]{1.00,1.00,1.00}{0}9&\textcolor[rgb]{1.00,1.00,1.00}{0}5&412396.2&0.9&412396.2&1.2\\
12&\textcolor[rgb]{1.00,1.00,1.00}{0}7&454917.7&1.6&454917.7&2.1\\
17&10&500752.3&1.3&500752.3&1.9\\
\hline\noalign{\smallskip}
\end{tabular}
\label{single_24RTS}
\end{table}

\subsection{IEEE 118-Bus Test System}
The second case study is based on the modified version of the IEEE 118-bus system examined in \cite{Minguez:16}. This test system comprises 118 buses, 54 generators, 91 loads, and 186 existing transmission lines, whereas a set of 61 candidate lines is available for expansion decisions \cite{data} under a \$100-million investment budget. In addition, a $\pm$50$\%$ fluctuation level was considered for all demands and generation capacities.

\begin{table}[t]
    \centering
    \caption{IEEE 118-Bus Test System -- Results}
\begin{tabular}{ @{\extracolsep{3pt}}c@{\extracolsep{4pt}}c@{\extracolsep{5pt}}c@{\extracolsep{3pt}}c@{\extracolsep{5pt}}c@{\extracolsep{3pt}}c }
\cline{3-6}\noalign{\smallskip}&&\multicolumn{2}{c}{P-CC}&\multicolumn{2}{c}{D-CC}\\
\hline\noalign{\smallskip}
%
&&Total&Computing&Total&Computing\\
$\Gamma^{\rm D}$&$\Gamma^{\rm G}$&cost&time&cost&time\\
&&(10$^6$ \$)&(s)&(10$^6$ \$)&(s)\\\hline\noalign{\smallskip}
10&\textcolor[rgb]{1.00,1.00,1.00}{0}5&19046.3&6.2&19279.2&27.4\\
20&15&23286.7&5.1&23353.4&\textcolor[rgb]{1.00,1.00,1.00}{0}9.8\\
60&35&30031.2&6.0&30033.9&25.1\\
91&54&31995.0&3.9&31995.0&12.0\\
\hline\noalign{\smallskip}
\end{tabular}
\label{single_118}
\end{table}

The superior performance of the proposed P-CC over D-CC is illustrated in Table \ref{single_118}. It is worth mentioning that the total costs of the best solutions attained by P-CC slightly differed from those found by D-CC by factors ranging between 0.0\% for the most conservative instance to $-$1.2\% for the least conservative solution, whereas the computational effort was significantly reduced by factors ranging between 77.4\%  for $\Gamma^{\rm D}=10$ and $\Gamma^{\rm G}=5$ and 48.0\% for $\Gamma^{\rm D}=20$ and $\Gamma^{\rm G}=15$.

\begin{figure}[t]
  \begin{center}
  \includegraphics[width=0.45\textwidth]{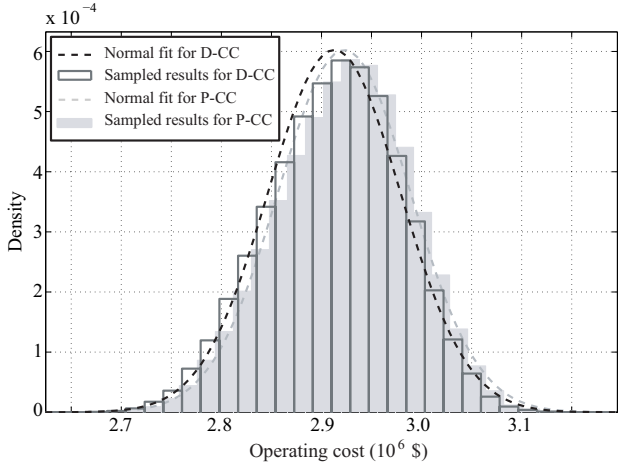}
    \caption{IEEE 118-bus test system -- Out-of-sample assessment for $\Gamma^{\rm D}=60$ and $\Gamma^{\rm G}=35$.}
    \label{Simulacion_35_60}
  \end{center}
\end{figure}

The quality of the best investment plans identified by both approaches has been further verified by an out-of-sample assessment based on the simulation of the operation of the expanded system for a random collection of different uncertainty realizations. To that end, the lower-level problem (\ref{of3})--(\ref{shed_lim}) has been solved for the corresponding best expansion plan and each sampled uncertainty realization. Due to space limitations, a representative instance, namely $\Gamma^{\rm D}=60$ and $\Gamma^{\rm G}=35$, has been selected to illustrate such an assessment. For this particular instance, out of the  $\binom {91} {60}\binom {54} {35}=3.63\times 10^{38}$ possible worst-case uncertainty realizations according to constraints (\ref{worstD})--(\ref{budgetG}), 100000 random samples were analyzed. Fig. \ref{Simulacion_35_60} shows the histograms and normal fitted densities of the operating cost for both methods. As can be seen, both distributions are very similar. Although the curves yielded by P-CC are displaced to the right, i.e., higher operating costs are incurred, the expected operating cost slightly increases by 0.39\%, which is acceptable bearing in mind that the computational performance is significantly improved by 76.1\%. Moreover, it is worth pointing out that no sampled operating cost exceeded the worst-case value for the best solution identified by P-CC, thereby substantiating the robustness of the proposed approach.

%


\subsection{Polish 2383-Bus Test System}
The third case study is based on the Polish system described in \cite{Polish}. This test system comprises 2383 buses, 327 generating units, 1500 loads, 2896 existing lines, and 124 candidate lines. Demand uncertainty is considered for the 100 largest loads, which are allowed to fluctuate by $\pm$20$\%$ around their nominal levels. Analogously, uncertainty in generation is modeled for the 212 generators with the lowest nominal generation capacities, for which their generation capacities are allowed to fluctuate between their nominal levels and zero. The interested reader is referred to \cite{data} for a full description of the benchmark.

Table \ref{single_2383} lists the results obtained for a \$100-million investment budget. As can be seen, for all uncertainty budgets examined, the proposed P-CC converged with acceptable computational effort for a planning setting. In contrast, for all those instances tested, D-CC was unable to find a single feasible solution after one week. It is worth stressing that the performance bottleneck of D-CC is the solution to the subproblem. As a matter of fact, for all simulations of this large-scale case study, D-CC exceeded the one-week time limit while attempting to solve the subproblem corresponding to the first iteration of the column-and-constraint generation algorithm. This result clearly evidences the computational gain associated with the use of the proposed coordinate descent method.

\begin{table}[t]
    \centering
    \caption{2383-Bus Test System -- Results}
\begin{tabular}{cccc} 
\hline\noalign{\smallskip}
%
%
\multirow{2}{*}{$\Gamma^{\rm D}$}&\multirow{2}{*}{$\Gamma^{\rm G}$}&Total cost&Computing time\\
&&(10$^6$ \$)&(s)\\\hline\noalign{\smallskip}
\textcolor[rgb]{1.00,1.00,1.00}{0}40&\textcolor[rgb]{1.00,1.00,1.00}{0}20&18823.4&42855.8\\
\textcolor[rgb]{1.00,1.00,1.00}{0}60&\textcolor[rgb]{1.00,1.00,1.00}{0}40&19514.2&11157.6\\
\textcolor[rgb]{1.00,1.00,1.00}{0}80&\textcolor[rgb]{1.00,1.00,1.00}{0}50&19591.7&\textcolor[rgb]{1.00,1.00,1.00}{0}3364.4\\
100&212&19689.9&\textcolor[rgb]{1.00,1.00,1.00}{0}3587.7\\
\hline\noalign{\smallskip}
\end{tabular}
\label{single_2383}
\end{table}

\begin{figure}[t]
  \begin{center}
  \includegraphics[width=0.45\textwidth]{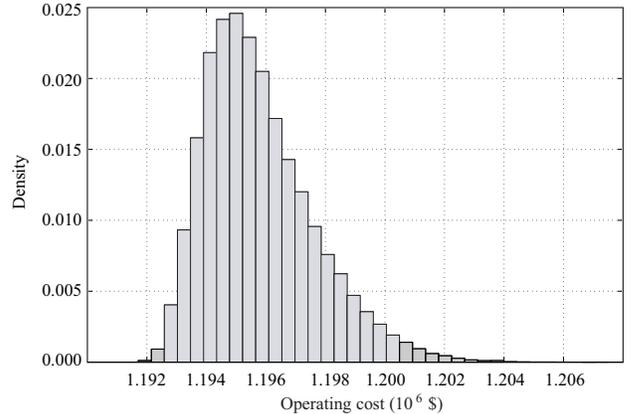}
    \caption{2383-bus test system -- Out-of-sample assessment for $\Gamma^{\rm D}=80$ and $\Gamma^{\rm G}=50$.}
    \label{Simulacion_50_80}
  \end{center}
\end{figure}

As done with the 118-bus system, the quality of the best investment plans provided by P-CC has been assessed via randomly sampling 100000 uncertainty realizations and computing the corresponding operating costs through the solution of the lower-level problem (\ref{of3})--(\ref{shed_lim}). For expository purposes, results from the out-of-sample assessment for $\Gamma^{\rm D}=80$ and $\Gamma^{\rm G}=50$ are reported. Fig. \ref{Simulacion_50_80} shows the histogram of the operating cost for this representative instance, for which $\binom {100} {80}\binom {212} {50}$ possible worst-case uncertainty realizations exist. For such uncertainty budgets, the maximum and the average sampled operating costs are equal to \$1.209 and \$1.196 million, respectively, which are far below the worst operating cost identified by the proposed method, namely \$2.236 million. These results confirm the robustness of the solution provided by the proposed P-CC.


\section{Conclusions}\label{Conclusions}
Existing methods relying on robust optimization are unable to solve large-scale instances of transmission network expansion planning under uncertain demand and generation capacity. This paper has presented a novel and computationally efficient primal column-and-constraint generation algorithm that overcomes such intractability issue. The strength of the proposed approach lies in its ability to address the subproblem associated with the max-min second-stage problem without resorting to the customary duality-based transformation to a single-level bilinear equivalent or its linearized version. To that end, a fast block coordinate descent algorithm is implemented on the space of primal decision variables, whereby two simple problems respectively associated with the middle- and lower-level problems are iteratively solved. As major advantages over previously reported methods, the proposed approach does not require duality-based cuts, a linearization scheme, or a case-dependent, non-trivial, and computationally expensive bounding parameter selection procedure for dual variables or Lagrange multipliers, which may be unbounded.

The effective performance of the proposed approach has been demonstrated with several case studies of different dimensions. For relatively small and medium-sized benchmarks, the proposed approach is able to identify expansion decisions that are either identical or very close to the best known solutions with substantially lower computational effort than that required by the state-of-the-art technique. Moreover, a practical test system with thousands of components is successfully handled by the proposed approach with moderate computing time, whereas the most efficient approach reported in the literature fails to provide a feasible solution within the allotted time limit.



\vspace{-1 cm}

\begin{IEEEbiography}[{\includegraphics[width=1in,height=1.25in,clip,keepaspectratio]{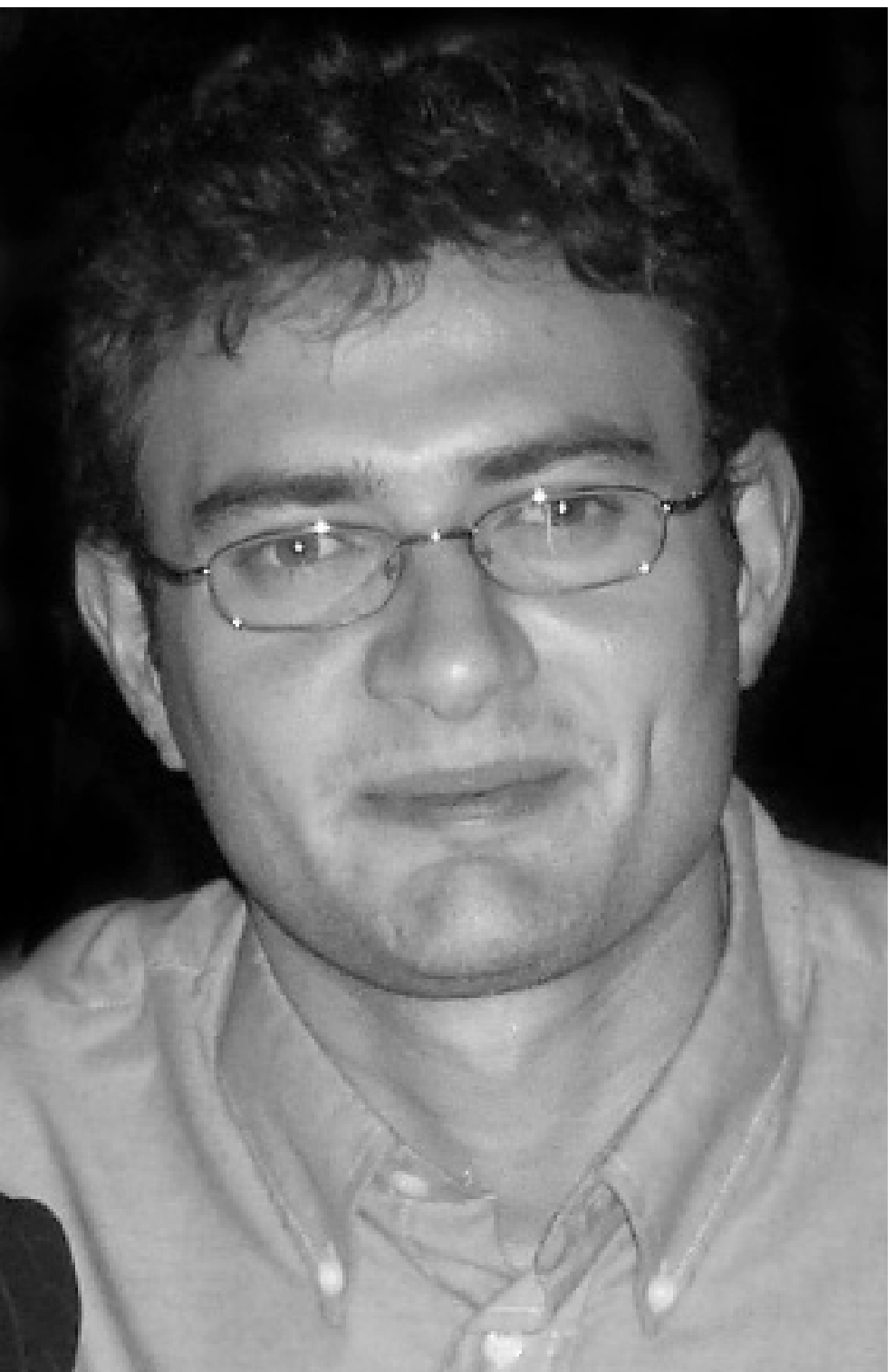}}]{Roberto M\'{\i}nguez} received the Civil Engineer degree and the Ph.D.
degree from the Universidad de Cantabria, Santander,
Spain, in 2000 and 2003, respectively.

He is currently a research fellow at the company Hidralab Ingeniería y Desarrollo, S.L., spin-off from the Universidad de Castilla-La Mancha. His research interests
include reliability engineering, sensitivity analysis, numerical methods, and optimization.
\end{IEEEbiography}
\vspace{-0.5 cm}
\begin{IEEEbiography}[{\includegraphics[width=1in,height=1.25in,clip,keepaspectratio]{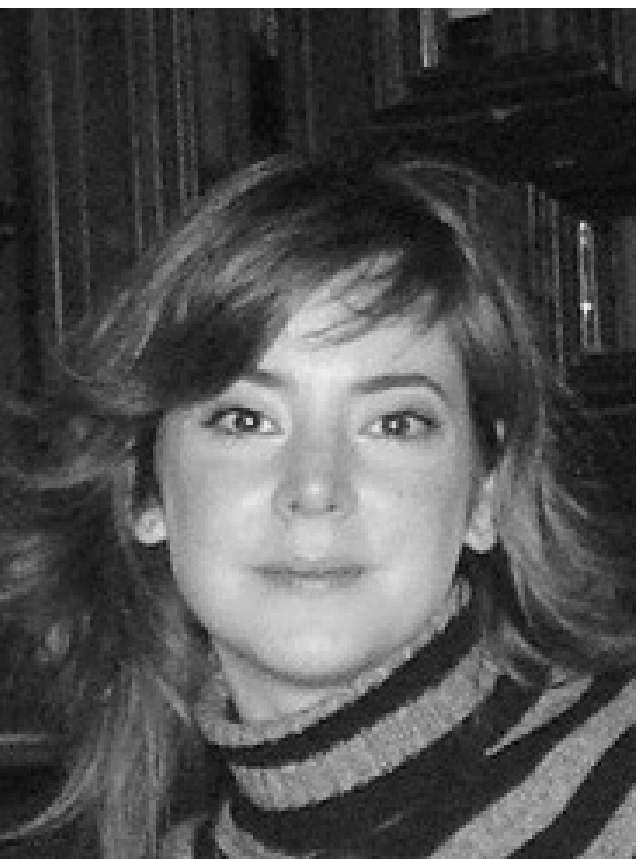}}]{Raquel García-Bertrand}
(S'02--M'06--SM'12) received the Ingeniera Industrial degree and the Ph.D.
degree from the Universidad de Castilla-La Mancha, Ciudad Real,
Spain, in 2001 and 2005, respectively.

She is currently an Associate Professor of electrical engineering at
the Universidad de Castilla-La Mancha. Her research interests
include operations, planning, and economics of electric
energy systems, as well as optimization and decomposition
techniques.
\end{IEEEbiography}
\vspace{-0.5 cm}
\begin{IEEEbiography}[{\includegraphics[width=1in,height=1.25in,clip,keepaspectratio]{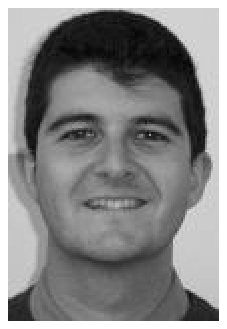}}]{José M. Arroyo}
(S'96--M'01--SM'06) received the Ingeniero Industrial degree from the
Universidad de Málaga, Málaga, Spain, in 1995, and the Ph.D. degree
in power systems operations planning from the Universidad de
Castilla-La Mancha, Ciudad Real, Spain, in 2000.

From June 2003 through July 2004 he held a Richard H. Tomlinson Postdoctoral Fellowship at the Department of Electrical and Computer Engineering of McGill University, Montreal, QC, Canada. He is
currently a Full Professor of electrical engineering at the
Universidad de Castilla-La Mancha. His research interests include
operations, planning, and economics of power systems, as well as
optimization.
\end{IEEEbiography}
\vspace{-0.5 cm}
\begin{IEEEbiography}[{\includegraphics[width=1in,height=1.25in,clip,keepaspectratio]{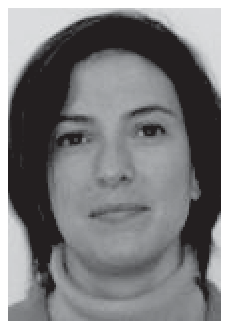}}]{Natalia Alguacil}
(S'97--M'01--SM'07) received the Ingeniero en Informática degree from
the Universidad de Málaga, Málaga, Spain, in 1995, and the Ph.D.
degree in power systems operations and planning from the Universidad
de Castilla-La Mancha, Ciudad Real, Spain, in 2001.

She is currently an Associate Professor of electrical engineering at
the Universidad de Castilla-La Mancha. Her research interests
include operations, planning, and economics of power systems, as
well as optimization.
\end{IEEEbiography}

\end{document}